\documentclass[aps,prd,eqsecnum,showpacs]{revtex4}
\usepackage{graphicx}
\begin{document}

\thispagestyle{empty}

\title{Criticality and convergence
in Newtonian collapse}
\author{Tomohiro Harada \footnote{Electronic
address:harada@gravity.phys.waseda.ac.jp},
Hideki Maeda \footnote{Electronic
address:hideki@gravity.phys.waseda.ac.jp} and
Benoit Semelin \footnote{Electronic
address:benoit.semelin@obspm.fr}}
\affiliation{Department of Physics, Waseda University, Shinjuku, Tokyo
169-8555, Japan}  
\date{\today}

\begin{abstract}                
We study through numerical simulation the spherical collapse of isothermal gas
in Newtonian gravity. We observe a critical behavior which occurs
at the threshold of gravitational instability leading to core formation. This was 
predicted in a previous work by two of the present authors. We describe it in
detail in this work.
For a given initial density profile, 
we find a critical temperature $T^{*}$, which is of the same order as the 
virial temperature of the initial configuration. For the exact critical 
temperature, the collapse converges to a self-similar form,
the first member in Hunter's family of self-similar solutions. 
For a temperature close to $T^{*}$, the collapse first approaches this 
critical solution. Later on,
in the supercritical case ($T < T^{*}$), the collapse converges to another 
self-similar 
solution, which is called the Larson-Penston solution. In the subcritical case
($T > T^{*}$),
the gas bounces and  disperses to infinity. We find two scaling 
laws with respect to  $|T-T^{*}|$:
one for the collapsed mass in the supercritical case 
and the other, which was not predicted before, for the maximum density reached 
before dispersal in the subcritical case.
The value of the critical exponent is measured to be $\simeq 0.11$ in the supercritical case, which agrees well with the predicted value $\simeq 0.10567$.
These critical properties are quite similar to those observed 
in the collapse of a radiation fluid in general relativity. 
We study the response of the system to temperature fluctuation and discuss
astrophysical implications for the interstellar medium structure and for the
star formation process.
Newtonian critical behavior is important not only because 
it provides a simple model for general relativity but also because it is 
relevant for astrophysical systems such as molecular clouds.
\end{abstract}

\maketitle
\section{introduction}
The Newtonian isothermal gas is a relevant description for 
important astrophysical
systems. It is indeed a good description for cold molecular clouds found in
the galactic interstellar medium. In these clouds the typical cooling time is
usually much shorter than the dynamical time; as a result they are in thermal
equilibrium with whatever steady heating source is present, for example 
the blackbody radiation of the cosmic microwave background, 
extragalactic ultraviolet radiation, or local stars. The 
molecular  clouds
can be described as isothermal. Since they are the prime location for star
formation, their physics determines the whole star formation history 
in the galaxy. A detailed understanding of the physics of these systems will 
lead to predictions for the properties of star formation.

There are two possible theoretical approaches for the self-gravitating 
Newtonian gas: statistical physics and dynamics. The traditional statistical
physics approach deals with equilibrium configurations of the system
and gave the first description of Jeans gravitational instability which is a 
fundamental behavior in the physics of self-gravitating 
systems (Chandrasekhar~\cite{chandrasekhar1939}, 
Padmanabhan~\cite{padmanabhan1990}).
Using the modern tools of statistical field theory and renormalization
groups, de Vega, Sanchez and Combes~\cite{devega1996} and Semelin 
{\sl et al.}~\cite{semelin1999} give a theoretical basis to  the observed
self-similar properties of the gas.
These statistical methods are useful, generally speaking, 
to describe systems undergoing second order phase transition and exhibiting 
critical behavior.
Investigating the critical behavior of the gas is 
also the subject of the present 
work, where we choose to describe the system through dynamics rather than
statistical physics and through the renormalization group.

The basic behavior of a clump of isothermal gas is the following.
In a sufficiently dense and/or cold configuration, the gas will 
collapse due to its self-gravity, while in a diffuse and/or hot one it will
expand due to its internal pressure forces.
A critical behavior appears at the transition between these two regimes.

In fact, Choptuik~\cite{choptuik1993} showed the existence
of a so-called critical phenomenon
in the spherical collapse of a massless scalar field
by general relativistic numerical simulations.
The simulations were performed for a 
one-parameter family of initial data sets.
He found a self-similar solution with critical properties 
at the threshold of black hole formation. In this first case, the 
self-similarity holds for a discrete set of transformations 
only (periodicity in logarithmic time).
He showed
that the mass of the formed black hole follows a scaling law,
and that the phenomenon is universal in the sense that 
it does not depend on the  one-parameter family of 
initial data sets. 
Soon after that, similar critical phenomena were observed in 
the spherical collapse of radiation fluid by 
Evans and Coleman~\cite{ec1994}.
In this second case, self-similarity holds for a continuous set of 
transformations.
Koike, Hara and Adachi~\cite{kha1995} gave a clear picture
of these phenomena by using the renormalization group approach.
Recently, critical phenomena have been found
in perfect fluids with more general 
equation of state~\cite{maison1996,kha1999,nc2000}
and in a variety of systems.
Astrophysical implications of the critical phenomena 
have been studied in the scenario of
primordial black hole formation by Yokoyama~\cite{yokoyama1998} 
and Niemeyer and Jedamzik~\cite{nj1998}. 
See \cite{gundlach1999} for a recent review of 
critical phenomena in gravitational collapse.

The solution with critical properties is only one of several known
self-similar solutions.
See Carr and Coley~\cite{cc1999} for a recent review of 
self-similar solutions in general relativity.
It was conjectured by Carr~\cite{carr1993} 
that a generic solution gradually evolves
toward a self-similar form, which suggest that 
one of the self-similar solutions may be an attractor solution. 
This is called the self-similarity hypothesis.
In Newtonian gravity, a self-similar solution,
the so-called Larson-Penston 
solution~\cite{penston1969,larson1969}
is known to be an attractor solution 
in the spherical collapse of isothermal 
gas~\cite{hn1997,hm2000,ti1999}.
In general relativity, Harada and Maeda~\cite{hm2001} showed 
by numerical simulations that 
the general relativistic counterpart of the 
Larson-Penston solution,
which was discovered by Ori and Piran~\cite{op1987},
is an attractor solution
in the spherical collapse of a perfect fluid with 
$P=k\epsilon $ ($0<k\alt 0.03$),
where $P$ and $\epsilon$ are the pressure and energy density,
respectively.
This convergence is quite important in general relativity
because it provides a strong counterexample against 
the cosmic censorship conjecture
in spherical collapse~\cite{op1987,op1990,harada1998,hm2001}. 
See \cite{hin2002} for a recent review of 
cosmic censorship and related 
topics. In the context of the self-similarity hypothesis, 
the stability criterion for general relativistic 
self-similar solutions against kink mode
was recently obtained by Harada~\cite{harada2001},
through a generalization of the Newtonian analysis~\cite{op1988}.

The present paper shows that critical phenomena 
exist in Newtonian gravity also.
We concentrate on a spherically symmetric system 
of isothermal gas in Newtonian gravity.
The existence of a critical phenomenon was predicted by Harada and 
Maeda~\cite{hm2001} and Maeda and Harada~\cite{mh2001}.
In these papers, based on the renormalization group analysis,
it was predicted that a critical 
phenomenon is associated with the first member of 
Hunter's family of continuous self-similar solutions~\cite{hunter1977}.
The criticality appears  in the scaling law for the collapsed mass, which is
satisfied for continuous values of the mass. The value of the critical 
exponent was predicted to be $\simeq 0.10567$.
The critical phenomenon observed in this paper
agrees very well with the above predictions, and gives a precise meaning to
the collapsed mass in Newtonian gravity.
We have also observed that the scaling law 
holds over many orders of magnitude
in terms of the parameter.

The organization of this paper is as follows.
In Sec. II, we apply the renormalization group analysis
to the Newtonian isothermal gas system
and draw the expected properties for the critical phenomenon.
In Sec. III, we present the results from
numerical simulations of Newtonian collapse.
In Sec. IV we study the response of the system to temperature fluctuations
and discuss astrophysical applications.
In Sec. V, we summarize the paper.

\section{Renormalization group analysis}
In this section, we review the renormalization group analysis
of the critical behavior in a spherical system of
Newtonian isothermal gas.
Part of the following discussion is based on
Maeda and Harada~\cite{mh2001}.

The dynamics of 
a spherically symmetric isothermal gas system
is described in the following set of equations:
\begin{eqnarray}
& &\frac{\partial\rho}{\partial t}+\frac{1}{r^2}
\frac{\partial}{\partial r}(r^2\rho v)=0, \label{b1}\\
& &\frac{\partial}{\partial t}(\rho v)
+\frac{1}{r^2}\frac{\partial}{\partial r}(r^2\rho v^2)
=-c_{\rm s}^2\frac{\partial\rho}{\partial r}-\rho\frac{G m}{r^2},  
\label{b2}\\
& &\frac{\partial m}{\partial t}+
v\frac{\partial m}{\partial r}=0,  \label{b3}\\
& &\frac{\partial m}{\partial r}=
4\pi r^2 \rho, \label{b4}
\end{eqnarray}
where $\rho=\rho(t,r)$, $v=v(t,r)$, $m=m(t,r)$, 
$c_{\rm s}$ and $G$
denote the density, 
radial velocity, total mass inside 
the radial coordinate, sound speed, 
and gravitational constant, respectively.
The sound speed $c_{\rm s}$ is related to the temperature $T$ of the gas as
\begin{equation}
c_{\rm s}=(kT/m)^{1/2},.
\end{equation}
where $k$ and $m$ are the Boltzmann constant and 
mass of each particle, respectively.

In order to apply the renormalization group
approach,
it is convenient to redefine
the physical quantities as follows:
\begin{equation}
\hat{r}\equiv \frac{r}{c_{\rm s}}, \quad
\hat{v}\equiv \frac{v}{c_{\rm s}}, \quad
\hat{m}\equiv \frac{m}{c_{\rm s}^{3}}. 
\end{equation}
In terms of these new quantities, the basic equations
reduce to the following form:
\begin{eqnarray}
& &\frac{\partial \rho }{\partial t}+\frac{1}{\hat{r}^2}
\frac{\partial}{\partial \hat{r}}
(\hat{r}^2 \rho  \hat{v})=0, \label{b11}\\
& &\frac{\partial}{\partial t}( \rho  \hat{v})
+\frac{1}{\hat{r}^2}\frac{\partial}{\partial \hat{r}}
(\hat{r}^2 \rho  \hat{v}^2)
=-\frac{\partial \rho }{\partial \hat{r}}
- \rho \frac{G\hat{m}}{\hat{r}^2},  
\label{b12}\\
& &\frac{\partial \hat{m}}{\partial t}+
\hat{v}\frac{\partial \hat{m}}
{\partial \hat{r}}=0,  \label{b13}\\
& &\frac{\partial \hat{m}}{\partial \hat{r}}=
4\pi \hat{r}^2  \rho . \label{b14}
\end{eqnarray}
We can see that the above hatted system 
does not contain any dependence on the temperature (or sound speed),
which simplifies the analysis.

The initial conditions for the bare system are 
specified by the density and velocity
profiles at the initial time $t=t_{\rm i} < 0$ 
($t=0$ is the time when the density singularity occurs 
in the center or, more physically,
the time of core formation).
Suppose the initial density and velocity profiles 
are given by functions $\rho_{\rm i}(r)$ 
and $v_{\rm i}(r)$, 
\begin{equation}
\rho(t_{\rm i},r)=\rho_{\rm i}(r) , \quad
v(t_{\rm i},r)=v_{\rm i}(r), 
\end{equation}
respectively.
Using the relations of the hatted quantities to the bare quantities,
we specify the initial conditions for the hatted system by 
\begin{equation}
\hat{\rho} (t_{\rm i},\hat{r})=\rho_{\rm i}(c_{\rm s}\hat{r}), \quad
\hat{v}(t_{\rm i}, \hat{r})=\frac{v_{\rm i}(c_{\rm s}\hat{r})}{c_{\rm s}}.
\end{equation}
Therefore, 
the parameter-dependent 
system~(\ref{b1})--(\ref{b4}) with unique initial conditions
is equivalent to the 
parameter-independent system~(\ref{b11})--(\ref{b14})
with a one-parameter family of initial conditions.
The parameter is  $c_{\rm s}$ or $T$.

Although we avoid to fully repeat the analysis 
by Koike, Hara and Adachi~\cite{kha1995},
we need to review it briefly.
We introduce the variables
\begin{equation}
\tau\equiv -\ln(-t), \quad
x\equiv  \ln\frac{\hat{r}}{-t}.
\end{equation}
$\tau$ is the scaling variable in the renormalization group
transformation.
Suppose a self-similar solution $H_{\rm ss}(x)$ exists,
with a unique relevant (unstable) mode, which turns out
to be the critical solution.
This self-similar solution can be 
considered as a fixed point of the renormalization group
transformation. 
The uniqueness of the unstable mode implies that 
the fixed point $H_{\rm ss}$ has a stable manifold
of codimension one.
This manifold is referred to as a critical surface. 
Any one-parameter family $\{H_{(p)}(x)|p\in 
{\bf R}\}$ of initial data sets with parameter $p$ 
generically has an intersection $H^{*}$ with 
the critical surface 
of the fixed point $H_{\rm ss}$.
$H^{*}$ will be driven to $H_{\rm ss}$ under the
renormalization group transformation, and hence
$H^{*}$ is the initial data set with the critical value $p^{*}$. 
We consider an initial data $H_{\rm i}(x)$ in the 
one-parameter family, close to $H^{*}$, i.e., 
\begin{equation}
h(0,x)=H_{\rm i}(x)=H^{*}(x)+\epsilon F(x),
\end{equation}
where $h(\tau,x)$ is a solution of 
the partially differential equations and $\epsilon=p-p^{*}$.
Then, for large $\tau_{0}$, we have
\begin{equation}
h(\tau_{0},x)\simeq H_{\rm ss}(x)+\epsilon e^{\kappa \tau_{0}}F_{\rm rel}(x),
\end{equation}
where $\kappa$ and $F_{\rm rel}(x)$ are the eigenvalue and eigenfunction
of the relevant mode, respectively.
Now we choose $\tau_{0}$ so that the first term and the 
second term are of the same order, i.e., 
\begin{equation}
\epsilon e^{\kappa \tau_{0}}=O(1).
\end{equation} 
At $\tau_{0}$, the solution is deviating
from the critical solution.
If the collapse bounces to expansion,
which will be the case for a subcritical collapse, 
the above condition obviously holds for the velocity field
at the moment of the bounce.
On the other hand, if the finite mass near the center 
shrinks considerably faster than for the critical collapse,
which will be the case for the supercritical case,
we have a $\tau_{0}$ for which the above condition also holds.
Since the deviation from the critical collapse
grows at $x=O(1)$,
the relation $\hat{r}=e^{x-\tau}$
implies that the radius of the core collapse or bounce
is given by 
$\hat{r}=O(e^{-\tau_{0}})=O(\epsilon^{\gamma})$, 
where $\gamma=1/(\mbox{Re}\kappa)$. 

In order to see clearly that self-similar solutions exist
in the present hatted system, we introduce the following quantities:
\begin{equation}
U\equiv\hat{v}, \quad
P\equiv 4\pi \hat{r}^{2} \rho, \quad
M\equiv \frac{G\hat{m}}{-t}.
\end{equation}
Then, the equations transform to
\begin{eqnarray}
& &-\dot{P}+(1+zU)P'+zPU'=0, 
\label{eq:basic1} \\
& &(\dot{U} P+ U\dot{P})-(U+zU^2+z)P' -(1+2zU)PU'-2zP+MPz^2=0,
\label{eq:basic2} \\
& &\dot{M}-M'-M+PU=0, 
\label{eq:basic3} \\
& &-z M'=P, 
\label{eq:basic4}
\end{eqnarray}
where the dot and prime denote the partial derivatives 
with respect to $\tau$ and $x$, respectively,
and $z\equiv e^{-x}$. 
The self-similar solutions are obtained by numerical integration
of the ordinary differential equations which 
follow from the above equations with the assumption
$P=P(x)$, $U=U(x)$ and $M=M(x)$.
If we assume the existence of the critical solution,
using the relation
\begin{equation}
m=\frac{c_{\rm s}^{3}}{G}M(x)e^{-x}\hat{r},
\end{equation}
we obtain the scaling law for the collapsed mass
or the ``core'' mass for the supercritical collapse, 
\begin{equation}
m_{\rm core}\propto|p-p^{*}|^{\gamma}.
\end{equation}
For the subcritical case, 
using the relation
\begin{equation}
\rho=\frac{P(x)}{4\pi}\hat{r}^{-2},
\end{equation}
we obtain the scaling law for 
the maximum density,
\begin{equation}
\rho_{\rm max}\propto |p-p^{*}|^{-2\gamma}.
\end{equation}

In fact, for Eqs.~(\ref{eq:basic1})--(\ref{eq:basic4}),
there is a discrete set of self-similar 
solutions with analyticity at the center and at the sonic point, 
such as the homogeneous free-fall, 
Larson-Penston,
Hunter (a), (b), (c) and (d) solutions, 
and so on.
Larson-Penston solution describes the coherent collapse
of the cloud for $t<0$.
This solution can be extended beyond $t=0$ 
to a late-time solution in which a finite mass 
collapses to the center and grows 
with time $t$.
Hunter (a) solution for $t<0$ describes an ``exploding collapse''
in which the mass of the central collapsing region gets smaller and 
smaller, converging to zero as $t$ reaches $0$, 
being surrounded by an expanding envelope.
The central density still blows up 
to a ``singularity'' at $t=0$.
This solution can also be extended beyond $t=0$
to a late-time solution in which the whole
cloud expands away;
the singularity disappears and 
the central density keeps decreasing.

Maeda and Harada~\cite{mh2001} found a unique
relevant (unstable) mode for the Hunter (a) solution but
no relevant mode for the Larson-Penston solution.
It was concluded that the former is a critical solution
while the latter is an attractor solution.
In the former case, the eigenvalue $\kappa$ 
of the unique relevant mode was evaluated to 
$\kappa\simeq 9.4637$ by solving the eigenvalue problem.
Other members of Hunter's family have more than
one relevant modes, which implies that these 
solutions are not critical solutions.
The homogeneous free-fall solution suffers from kink 
instability~\cite{op1988}.
The value of the critical exponent $\gamma$ is $\simeq 0.10567$.
Since we have seen that the parameter for the family 
of initial data sets is given by the temperature $T$, 
the scaling laws in the present system are given by
\begin{equation}
m_{\rm core} \propto |T-T^{*}|^{\gamma}
\end{equation}
for the supercritical case, and 
\begin{equation}
\rho_{\rm max}\propto |T-T^{*}|^{-2\gamma},
\end{equation}
for the subcritical case.
It should be noted that the parameter is not necessarily 
the temperature but may be 
the mass or the radius of the cloud
if we fix the temperature.

\section{numerical simulations}
We have performed numerical simulations of the spherical collapse
of an isothermal Newtonian gas.
The basic equations for the evolution of the system are given by
Eqs.~(\ref{b1})--(\ref{b4}). 
The code is based on the Lagrangian description of hydrodynamics and 
a finite difference scheme.
We have chosen Lagrangian (comoving) radial coordinates, attached to 
spherical shells with infinitesimal thickness. The mass inside
coordinate $\bar{r}$
is consequently constant throughout the simulation. We call such coordinates
mass coordinates. Shells are initially positioned at constant intershell
distance and their mass is computed from the desired density profile (we do
not use equal mass shells).
With these choices, we achieve high accuracy for the dynamics in
the dense central region where the shells collapse during the simulations. 
The number of shells used in the simulation is $10^4$. One simulation with 
$10^5$ shells has been run to check the influence of numerical resolution.
No effect has been detected. This is proof of the accuracy of the code.
In this section, we adopt units in which $G=k/m=1$ and
the total mass $m_{\rm tot}$ and 
initial radius $r_{\rm surf,i}$ of the cloud are unity.

We have computed the evolution of four types of initial density profiles.  
Details of the models for the initial density profiles 
are summarized in Table~\ref{tb:Tc}.
The initial velocity profile is always set to be identically zero.
This assumption seems the most relevant for astrophysical applications, but
the same dynamics should develop for different initial velocity profiles.
Model 2 is our reference model, in which
the initial density profile is given by a quartic polynomial
and both the density and the density gradient (i.e., pressure forces) are 
continuous at the cloud surface.
We have computed collapse simulations for various values 
of the temperature.
(Actually, we have computed for various sound speeds and 
calculated the temperature because the sound speed is 
rather dynamically meaningful.)

For each model, we find a critical temperature $T^{*}$, which is of the same order of magnitude as the
virial temperature estimated from the initial gravitational energy of the 
system. 
For temperatures just below $T^{*}$, the cloud collapses and 
the central density goes to infinity in a finite time, producing a singularity.
We call this collapse a supercritical collapse. 
For temperatures just above the critical 
temperature,
the cloud begins to collapse but bounces afterward, and the density finally 
goes to zero everywhere. We call this incomplete ``collapse'' a subcritical collapse.
The results for the critical temperture for each model are
summarized in Table~\ref{tb:Tc}.
In this table, we quote only the number of digits that are 
common between the 
critical values computed using the two different resolutions,
$10^{4}$ and $3\times 10^{4}$.

\begin{figure}[htbp]
\rotatebox{90}{
\includegraphics[scale=1]{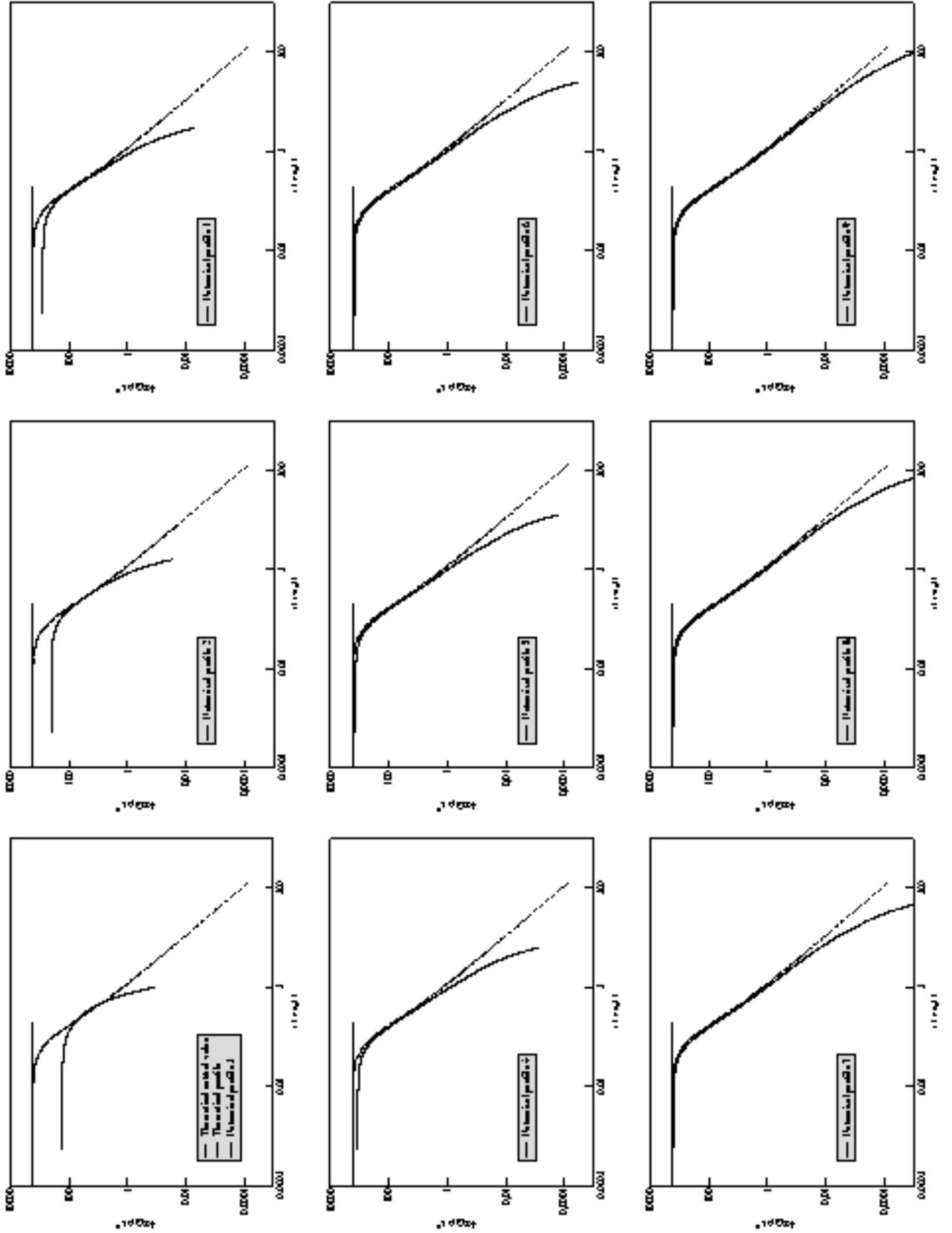}}
\caption{\label{fg:hunter}
Snapshots of density profile are plotted in terms of the dimensionless quantity
$4\pi G \rho t^2$ as a function of $r/(-c_{\rm s} t)$, for the critical value
of the temperature in model 2 ($|\delta T|/T^{*}\equiv 
|T-T^{*}|/T^{*}\sim 10^{-15}$) at different times. 
The origin of time coordinate is set to be 
the time of singularity formation.
The theoretical Hunter (a)
solution is also plotted. Numerical profiles converge to the 
Hunter (a) solution.}
\end{figure}

In the critical and supercritical cases,
we can estimate the time $t_{\rm sing}$ of singularity formation
at the center in different ways. First we can just define it as the instant
when the numerical
density reaches a very high threshold (e.g., $\rho \sim  10^{15}$ 
while initially
of the order of one). Or we can assume the convergence to a self-similar 
behavior and use the relation $\rho_{\rm c}\propto (t_{\rm sing}-t)^{-2}$
for the central density $\rho_{\rm c}$. Both approaches have been tried and
give the same results. 
When the temperature is fine-tuned to the critical one, a case that we call 
the critical case for convenience, the simulations show that
the collapse approaches the Hunter (a) solution 
for a while before it deviates. The finer the tuning of the temperature, the
closer the approach and the later the deviation.
Figure~\ref{fg:hunter} shows  density profiles in terms of the dimensionless quantity
$4\pi \rho (t_{\rm sing}-t)^{2}$ for a tuning of the temperature to the critical
temperature $|\delta T| / T^{*} \equiv |T-T^{*}|/T^{*}
\sim 10^{-15}$, along with the theoretical
Hunter (a) solution. In this case we estimate $t_{\rm sing}$ using the
$\rho_{\rm c}\propto (t_{\rm sing}-t)^{-2}$ relation. This estimation is local
in time and it gives 
the value which $t_{\rm sing}$ would take if Hunter (a) solution was
fully realized.
We can see that the density profile converges to Hunter 
(a) solution. 
A discrepancy appears at large radii 
because the numerical configuration has a finite
mass while the theoretical Hunter (a) solution has an infinite mass. This discrepancy
moves to larger and larger value of $r/(-c_{\rm s}t)$ as we approach $t_{\rm sing}$.
Therefore, we associate the Hunter (a) solution to the critical case.

\begin{figure}[htbp]
\includegraphics[scale=0.6]{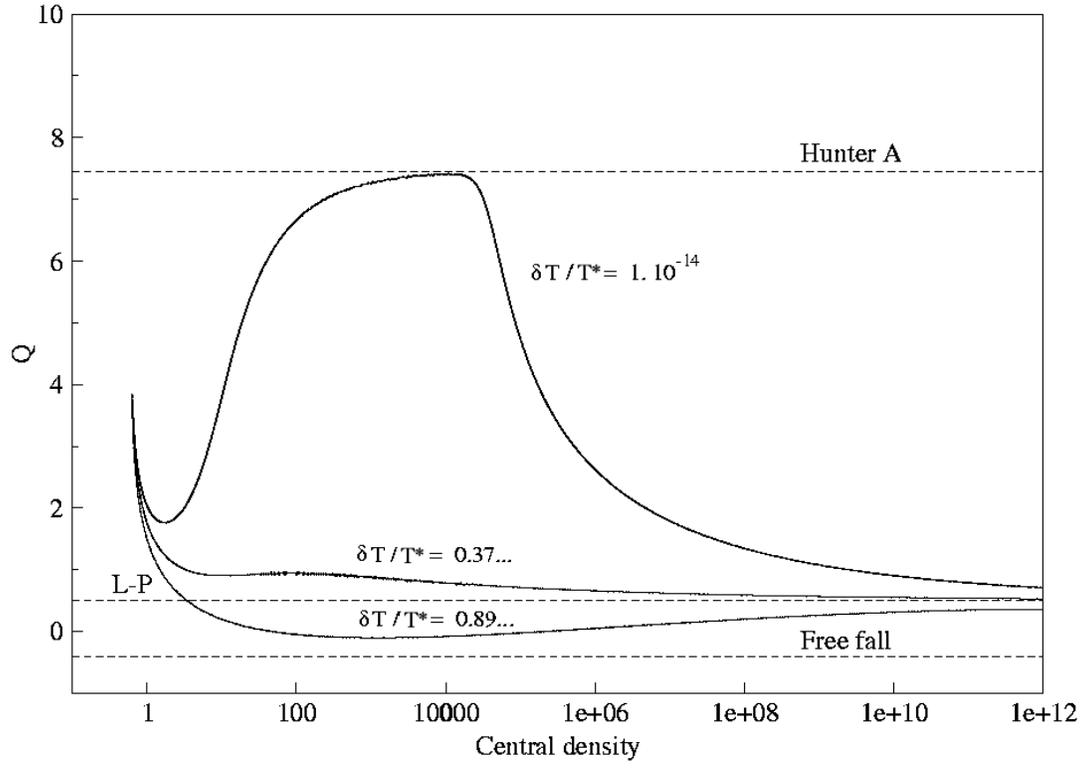}
\caption{\label{fg:Q}
Evolution of $Q\equiv \ln [4\pi G\rho_{\rm c} (t_{\rm sing}-t)^{2}]$
as a function of the central density $\rho_{\rm c}$, for model 2.
The time $t_{\rm sing}$ of singularity formation is estimated locally.
The theoretical values of $Q$ for several self-similar solutions are 
also plotted in this figure.
See text for details.}

\end{figure}

\begin{figure}[htbp]
\rotatebox{90}{
\includegraphics[scale=1]{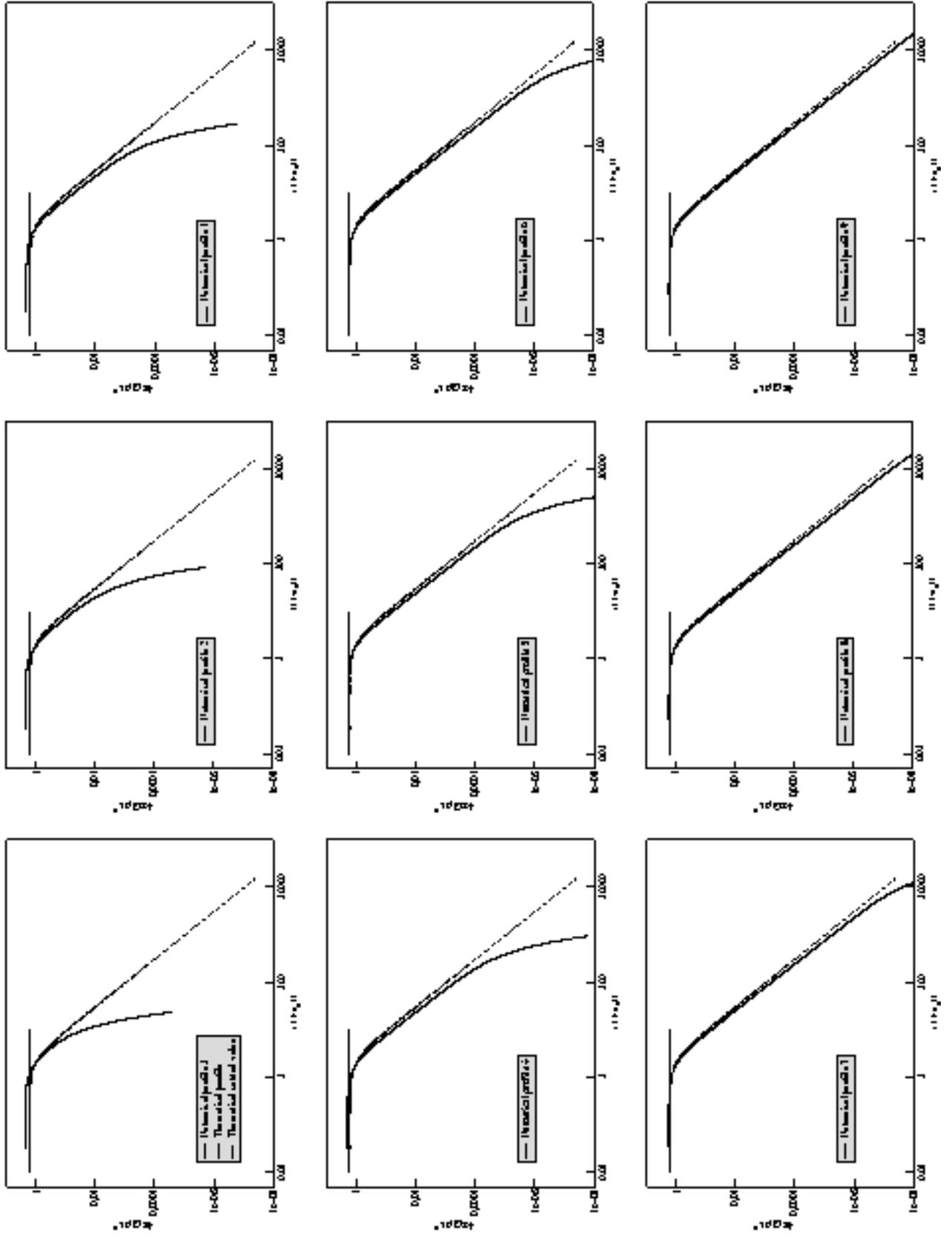}}
\caption{\label{fg:larson}
Snapshots of density profile are plotted in terms of the dimensionless quantity
$4\pi G \rho t^2$ as a function of $r / (-c_{\rm s} t)$, for a supercritical
value of the temperature in model 2 ($\delta T /T^{*}=-1/2$)
at different times. 
The origin of time coordinate is set to be 
the time of singularity formation.
The theoretical Larson-Penston
solution is also plotted. Numerical profiles converge to the Larson-Penston
solution.}
\end{figure}

In the near-critical supercritical case, 
the collapse first approaches the critical solution.
Figure~\ref{fg:Q} shows the evolution of the nondimensional 
central density parameter
\begin{equation}
Q\equiv \ln [4\pi G\rho_{\rm c} (t_{\rm sing}-t)^{2}],
\end{equation}
as a function of the central density $\rho_{\rm c}$ itself.
Here again, we estimate $t_{\rm sing}$ locally. We compute 
$t_{{\rm sing},n}$ at each time step $n$,  
using the following relation
\begin{equation}
\rho_{{\rm c},n}(t_{{\rm sing},n}-t_{n})^{2}
=\rho_{c,n-1}(t_{{\rm sing},n}-t_{n-1})^{2},
\end{equation}
where $t_{n}$ and $\rho_{{\rm c},n}$ are the time and 
central density at the $n$-th step.
We can then deduce $Q$.
In this figure, the theoretical values for self-similar solutions,
such as the homogeneous free-fall,
Larson-Penston and Hunter (a) solutions,
are also plotted.
We find that the curve of the critical collapse
approaches the value of the Hunter (a) (critical) solution.
However, it afterward deviates and approaches the value of 
the Larson-Penston solution. This happens because the tuning of the temperature
to the critical temperature is not perfect, and because the critical solution
is unstable against such fluctuations. 
During and after the approach to the critical solution, the surrounding
region expands away and leaves a finite-mass ``core''
in the central region.
The density field 
in the central collapsed core at late times is well 
described by the Larson-Penston solution. This is the case early on for a 
supercritical collapse also, where the temperature is not fine-tuned to the 
critical temperature. This type of collapse also produce a finite (more massive)
core and an expanding shell.
We can see the convergence to Larson-Penston solution in Fig.~\ref{fg:larson}.
Therefore we identify the Larson-Penston solution
as an attractor solution.

\begin{figure}[htbp]
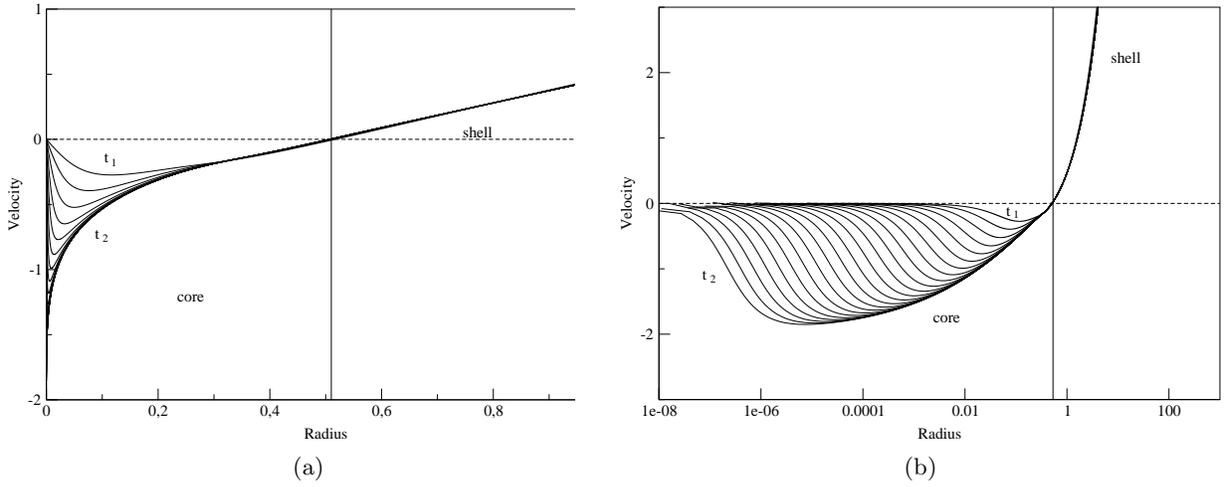

\begin{center}
\begin{tabular}{cc}
\includegraphics[width=8cm]{4a} &
\includegraphics[width=8cm]{4b}\\
(a) & (b) \\
\end{tabular}
\caption{\label{fg:core}
The evolution of the velocity profile for a supercritical collapse
(model 2), (a) in linear scale and (b) in logarithmic scale.
The temperature is about 1\% less than the critical value.
A collapsed core forms in the center, surrounded by an expanding shell.
Label $t_{1}$ is the time  when the central density reaches 10, and
label $t_{2}$ is the 
time when we have stopped the calculation.}  
\end{center}

\end{figure}

\begin{figure}[htbp]
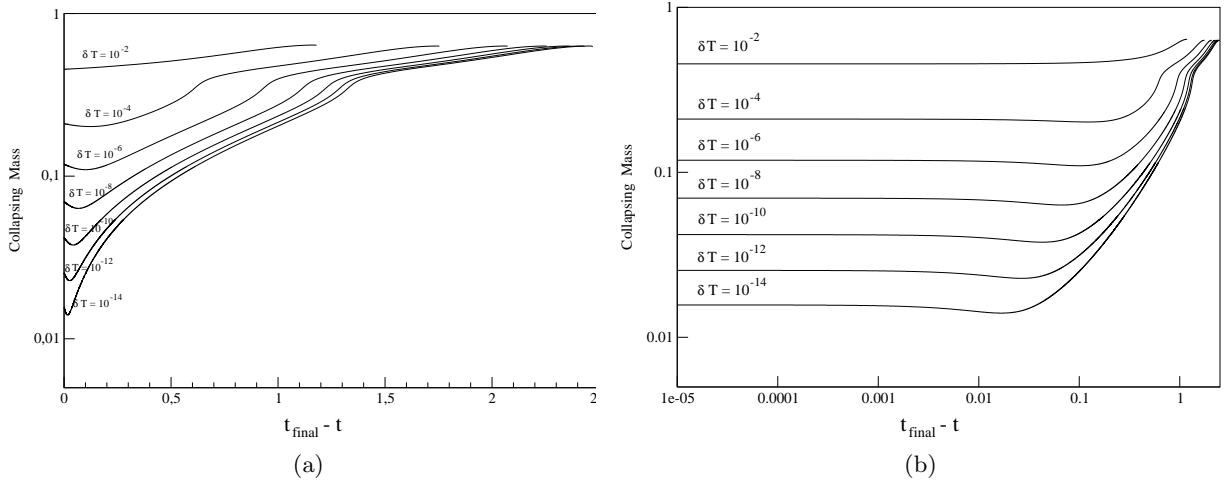
 
\begin{center}
\begin{tabular}{cc}
\includegraphics[width=8cm]{5a} &
\includegraphics[width=8cm]{5b}\\
(a) & (b) \\
\end{tabular}
\caption{\label{fg:coremass} Evolution of the collapsing mass 
as a function of
time for different values of supercritical temperatures 
[($\delta T=T-T^{*}$) 
difference to the critical temperature] for model 2, (a) in linear scale 
and (b) in logarithmic scale. 
The collapsing mass
is defined as the total mass of shells having a negative radial velocity. For
all values of the temperature the mass reaches a constant value early on, 
forming a well defined ``core''.}
\end{center}
\end{figure}

\begin{figure}[htbp]
\includegraphics[scale=0.6]{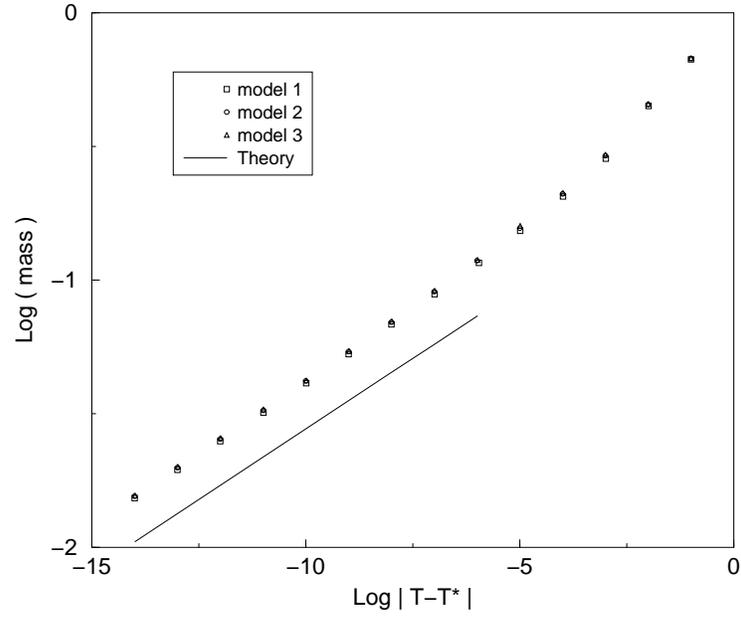}
\caption{\label{fg:massscale} Scaling law for the mass of the collapsed core
in supercritical collapses.
The masses of the collapsed core formed in supercritical collapses
in each model are plotted for different temperatures
just below the critical temperature.
The theoretical slope determined by the renormalization group analysis
is also depicted.}
\end{figure}

Figure~\ref{fg:core} shows the velocity profile around the center for a supercritical collapse.
We can see that a finite-size collapsing core is formed.
Moreover the 
physical radius of the collapsing core is almost constant.
Figure~\ref{fg:coremass} 
shows the time evolution of the collapsing mass, for different 
(supercritical) values of the temperature.
We can see that at all supercritical temperature, the collapsing mass tends 
to a constant and therefore the ``core'' mass is well defined 
at the time of the singularity.
Figure~\ref{fg:massscale} shows the relation between 
the core mass and the temperature fluctuation $|T-T^{*}|$.
The numerical results reproduce the scaling law derived by the renormalization
group analysis in the previous section.
The numerical value of the critical exponent is $\simeq 0.11$, which agrees well
with the theoretical value $\simeq 0.10567$
determined from the eigenvalue problem
of the linear mode analysis.
It should be noted that the power-law relation
holds approximately
for $|\delta T|/T^{*}\alt 0.01$.

\begin{figure}[htbp]
\includegraphics[scale=0.6]{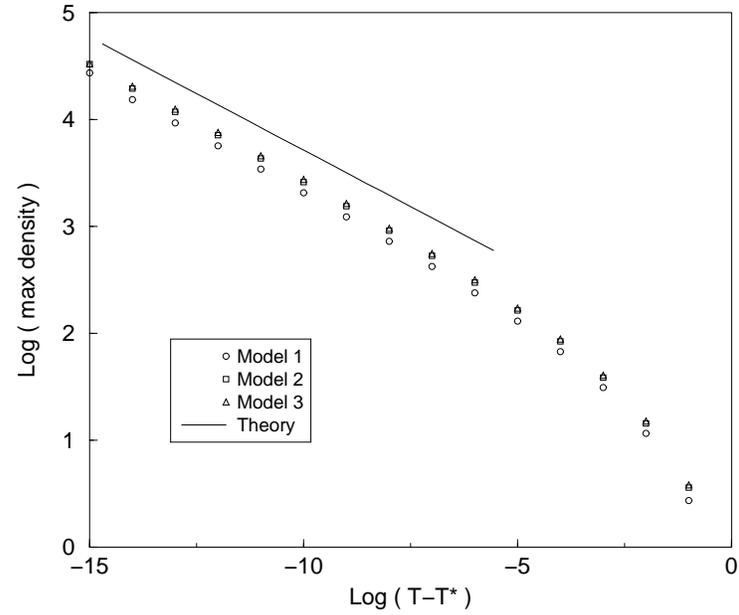}
\caption{\label{fg:densscale} Scaling law for the maximum density reached
in subcritical collapses.
The maximum densities reached in subcritical collapses
in each model are plotted for different temperatures
just above the critical one.
The theoretical slope determined by the renormalization group analysis
is also depicted.}
\end{figure}

For the subcritical collapse, the central density reaches a maximum 
and then bounces away.
Figure~\ref{fg:densscale} shows the relation between
the maximum density and the temperature excess $T-T^{*}$.
It shows that the maximum density also follows 
a scaling law, as is expected from the renormalization group 
analysis. The measured value $\simeq -0.22$ 
of the critical exponent agrees with 
the theoretical value $\simeq -0.21134$.
Again, it should be noted that the power-law relation
holds approximately for $\delta T/T^{*}\alt 0.01$.

Although we have concentrated on model 2, 
the above features have been checked for models 1 and 3. They are identical in
all three models except for the specific value of the critical temperature.
Model 4, the top-hat model, is special
because the cloud is initially homogeneous.
In this case, the central region of the cloud 
remains homogeneous and freely falls to a singularity
if the condition $m_{\rm tot}/r_{\rm surf,i}> \pi^{2}c_{\rm s}^{2}/2$
is satisfied~\cite{ti1999}.
If the condition is not satisfied,
the collapse converges to Larson-Penston
(attractor) solution.

\section{astrophysical implication}
In this section, we study 
the mass spectrum of  collapsed cores resulting from temperature fluctuations
around the critical case, according to the 
treatment of Yokoyama~\cite{yokoyama1998}.
We build a toy model which shows how the critical gravitational collapse
may affect the properties of the interstellar medium, and in particular the 
processes of fragmentation in molecular clouds and star formation.
It should be emphasized that the purpose of this section is 
not to derive the whole shape of the observed initial mass function (IMF)
but to show how a link can be established between an initial fluctuation field,
and the resulting mass spectrum for the collapsed objects. Numerous additional
physical processes would have to be introduced to derive an IMF.

We will use 
\begin{equation}
M_{\rm core}(T) = K M_{\rm tot}\left(1- \frac{T}{T^{*}} \right)^\gamma,
\label{eq:massscalinglaw}
\end{equation}
for the masses of cores formed by collapsing isothermal gas clouds with mass 
$M_{\rm tot}$ and a temperature which is slightly less than $T^{*}$. $K$ is
a constant and $\gamma \approx 0.11$ is independent of initial density 
profile. This relation is theoretically valid for $\mid T-T^{*}\mid \ll T^{*}$.
We can check on Fig.~\ref{fg:massscale} that it is actually valid for
$|T-T^{*}|/T^{*} \alt 0.01 $ or $M_{\rm core} \alt 0.3 M_{\rm tot}$.

We assume a Gaussian probability distribution for the temperature 
fluctuations for simplicity, 

\begin{equation}
\label{pt}
P(T) = \frac{1}{\sqrt{2\pi} \sigma} \, \exp{\left(-\frac{(T-T_{\rm m})^2}{2\sigma^2}\right)},
\end{equation}
where $T_{\rm m}$ is the temperature with the maximum probability and
$\sigma$ is the dispersion, which has the dimension of a temperature. 
For simplicity, we consider a collection of clouds of the 
same mass and size with a
distribution of temperature obeying the above probability function.
Equation (\ref{pt}) allows us to compute the
fraction of clouds collapsing to cores
\begin{equation}
\beta = \int^{T^{*}}_{T^{*}(1-K^{-1/\gamma})} P(T) dT.
\end{equation}
The lower integration limit reflects the fact that 
a core with a mass heavier than the initial gas cloud cannot be formed.

We can now determine the mass sprectrum of cores  formed from a collection 
of gas clouds with masses $M_{\rm tot}$. We define the mass spectrum as the
number $dN$ of cores per logarithmic mass bin, normalized by 

\begin{equation}
\label{norm}  
\int_{-\infty}^{{\rm ln}M_{\rm tot}}\frac{dN}{d({\rm ln}M_{\rm core})}d({\rm ln}M_{\rm core}) = 1.
\end{equation}
This mass function is given by

\begin{eqnarray}
\label{phi}
\frac{dN}{d({\rm ln}M_{\rm core})}& = & -\beta^{-1}P(T (M_{\rm core}))
\frac{dT}{d({\rm ln}M_{\rm core})}, \nonumber \\
&= &\frac{T^{*}}{\sqrt{2\pi} \beta\sigma\gamma} \left(\frac{M_{\rm core}}{K M_{\rm tot}}\right)^{1/\gamma}{\rm
exp}\left(-\frac{1}{2\sigma^2}\left[T^{*}\left\{1-\left(\frac{M_{\rm core}}{K M_{\rm tot}}\right)^{1/\gamma}\right\}-T_{\rm m}\right]^2\right).
\end{eqnarray}

\begin{figure}[htbp]
\begin{center}
\begin{tabular}{cc}
\includegraphics[width=8cm]{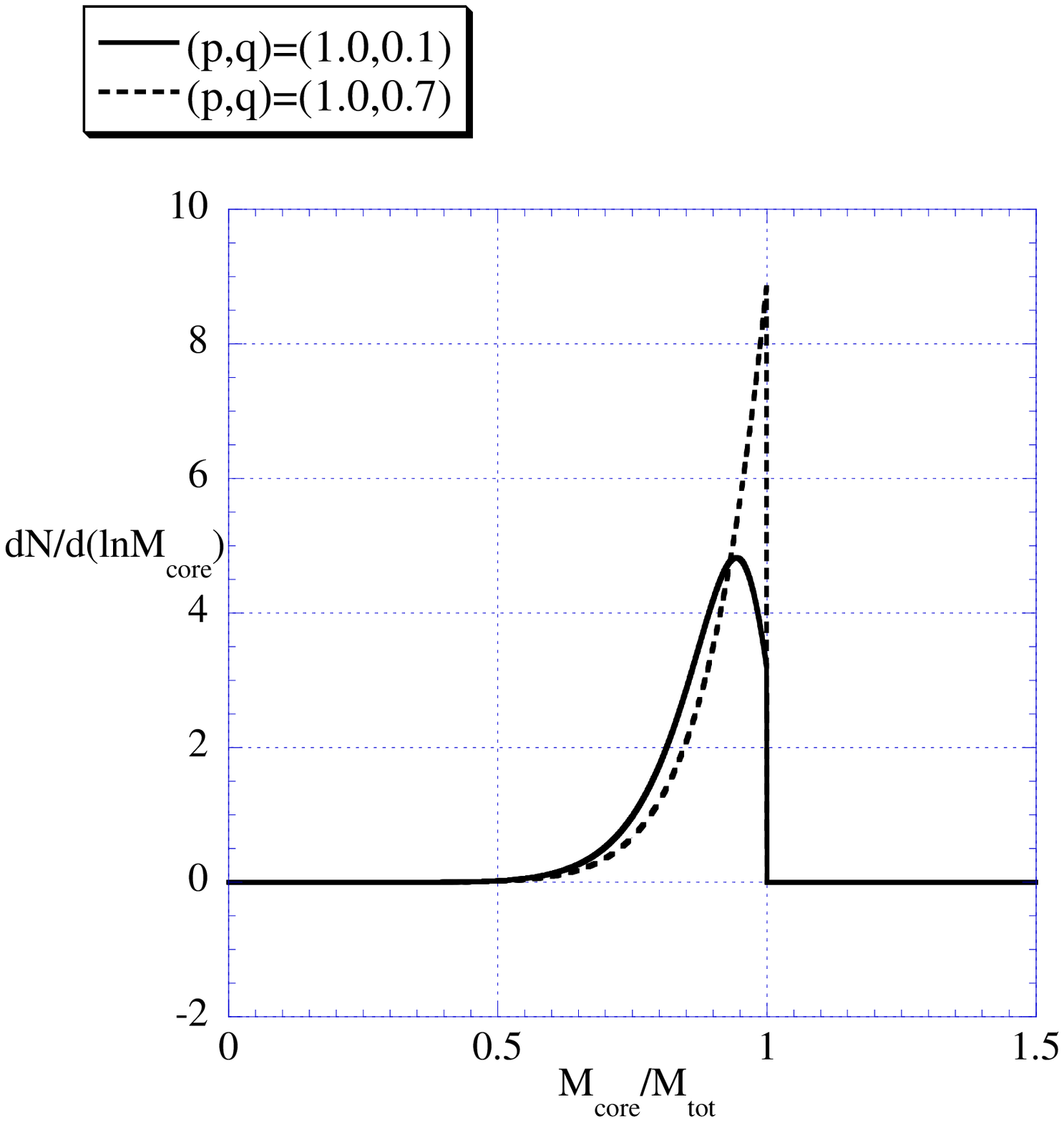}&
\includegraphics[width=8cm]{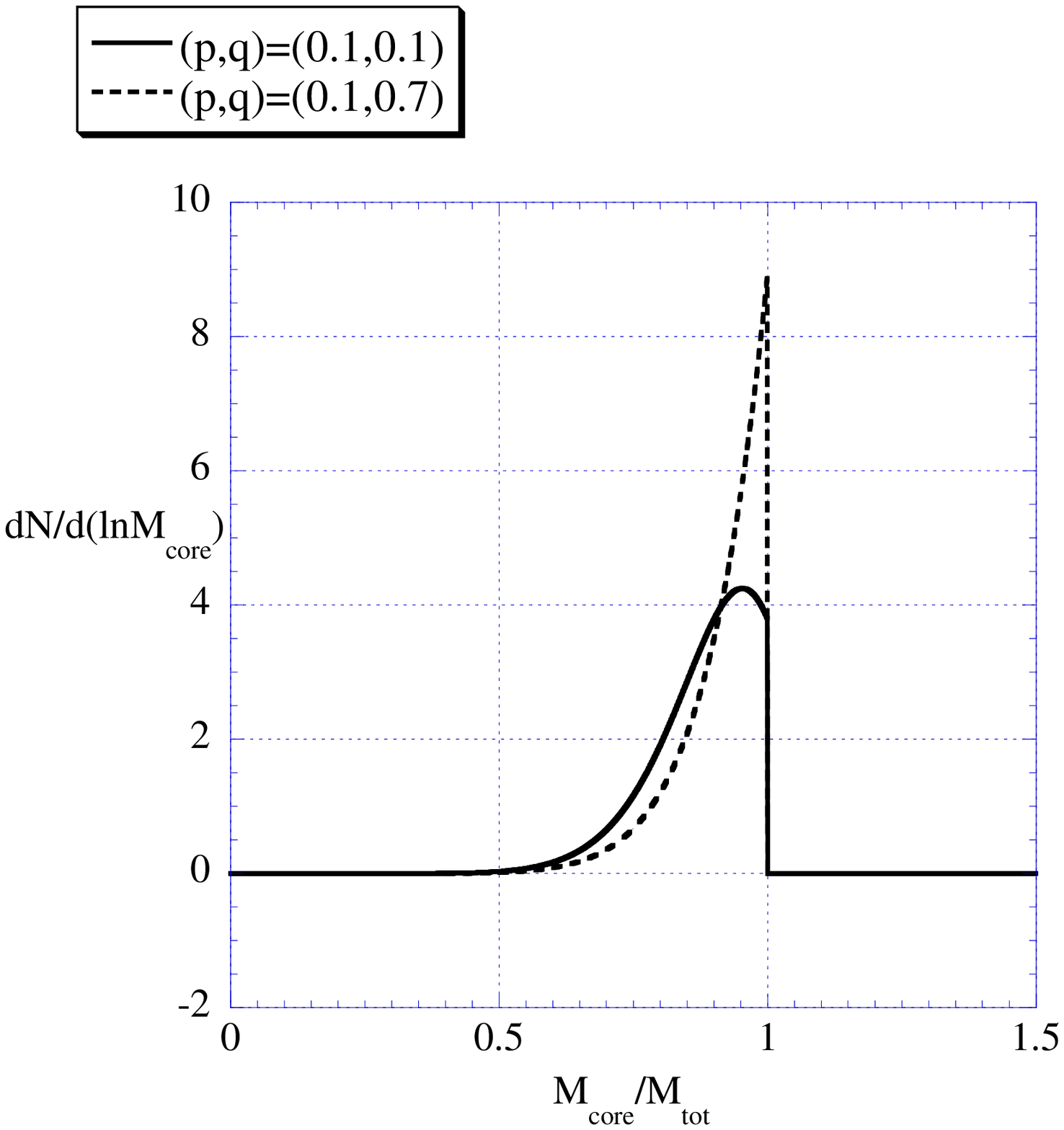}\\
(a)& (b) \\
\end{tabular}
\caption{\label{fg:imf} Mass functions with four sets of
parameter values for $p \equiv T^{*}/T_{\rm m}$ and 
$q \equiv \sigma/T_{\rm m}$ for
 $\gamma = 0.11$ and $K = 1.2$. 
In (a), the solid and the dashed lines
correspond to $(p,q)=(1.0,0.1)$ and $(p,q)=(1.0,0.7)$, respectively.
In (b), the solid and the dashed lines correspond 
to $(p,q)=(0.1,0.1)$ and $(p,q)=(0.1,0.7)$, respectively.}
\end{center}
\end{figure}

From numerical simulations, we have obtained $\gamma \sim 0.11$ and 
$K \sim 1.2 $. 
There are two parameters $p \equiv T^{*}/T_{\rm m}$ and $q \equiv \sigma/T_{\rm m}$ in
the mass spectrum. The mass spectra are depicted for $(p,q)=(1.0,0.1)$ and
$(p,q)=(1.0,0.7)$ in Fig.~\ref{fg:imf}(a), while for $(p,q)=(0.1,0.1)$
and $(p,q)=(0.1,0.7)$ in Fig.~\ref{fg:imf}(b). Core masses heavier
than the total mass are cut off. The mass $M_p$ associated with the peak 
in the spectra is smaller than $M_{\rm core}=M_{\rm tot}$ for $(p,q)=(1.0,0.1)$ 
and $(0.1,0.1)$, while equal to $M_{\rm core}=M_{\rm tot}$ for
$(p,q)=(1.0,0.7)$ and $(0.1,0.7)$ [these regions are however not fully relevant
because of the restrictions to be applied to
Eq.~(\ref{eq:massscalinglaw}]. 
$M_{\rm p}$ and
the average mass $M_{\rm av}$ of the formed core defined as

\begin{equation} 
M_{\rm av} \equiv \frac{\int^{M_{\rm tot}}_{0}M_{\rm core}\frac{dN}{d M_{\rm core}}d M_{\rm core}}{\int^{M_{\rm tot}}_{0}\frac{dN}{d M_{\rm core}}d M_{\rm core}},
\end{equation}
are summarized in Table \ref{tb:IMF}. Although $M_{\rm p}$ can be just 
equal to $M_{\rm tot}$, 
the average mass $M_{\rm av}$ is always smaller. 

The main feature of the mass spectrum is a steep slope in
the low-mass tail, due to the small value of the critical exponent $\gamma$.
This fact has several implications.  First, the cores are created with nearly 
the mass of the 
initial gas cloud. In the formalism originally introduced by Press and 
Schechter~\cite{ps1974}, it is assumed that overdense regions with a single 
mass scale, if they collapse, result in bound objects with the same mass. 
However, our 
result implies that the mass of the collapsed object is not the same as 
the mass of the initial overdense region, but may be rather smaller if the 
 collapse is near-critical. 
 Second, the critical collapse alone cannot account for the shape
 of the observed IMF, which shows a {\em negative} exponent on most of the
 mass spectum while we find a large {\em positive} exponent. Other phenomenon 
 are responsible for the negative exponent.

\section{summary}
We have analyzed the critical behavior in the collapse of a Newtonian isothermal
gas by the renormalization group approach and in numerical simulations.
The renormalization group analysis in the Newtonian case is similar to the 
general relativity case.
We have observed the critical phenomenon in the numerical simulation,
which can be compared to those
observed during black hole formation 
in the collapse of radiation fluid in general relativity.
The scaling law for the formed
black hole mass with critical exponent $\simeq 0.36$ in general relativity
is replaced by a scaling law for the collapsed core mass
with critical exponent $\simeq 0.11$.
The critical solution is a continuous self-similar solution,
specifically the first member of Hunter's family of solutions.

For the supercritical collapse,
the collapse near the center converges (early for low temperature, late for
almost critical temperature) to 
another continuous self-similar solution, the Larson-Penston solution.
Generic collapse converges to this attractor solution,
which suggests the universality of self-similarity 
in gravitational collapse.
This phenomenon is an example for the validity of 
the self-similarity hypothesis in gravitational collapse.

Some implications of Newtonian critical phenomena
to astrophysics have been discussed.
The critical behavior alone does not determine
the IMF, clumps mass spectrum in molecular clouds and so on.
However it suggest some limitations to Press-Schechter's model.
The isothermal gas is a good approximation for a gas system with 
a short cooling time, while
it will not be valid for optically thick clouds,
where adiabatic gas model are more relevant.
Consequently, critical phenomena should be investigated for more general 
equations of state.
Research along this line is now under investigation~\cite{maeda2002}. 

Newtonian gravity is a simple 
model for more complicated gravitational theories, 
such as general relativity.
Newtonian critical behaviors in gravitational collapse
provide a good laboratory for investigating various effects
on the critical phenomena in general relativity,
such as deviation from spherical symmetry, rotation,
external environment, a variety of matter fields,
and so on.

\acknowledgments

We are grateful to T.~Koike, B.J.~Carr and R.~Tavakol 
for helpful comments.
TH and BS were supported by the 
Grant-in-Aid for Scientific Research (Nos. 05540 and 00273)
from the Japanese Ministry of
Education, Culture, Sports, Science and Technology.

\newpage 

\begin{table}[htbp]
  \caption{\label{tb:Tc} Models of initial density profiles.}
\begin{center}
    \begin{tabular}{c|cc} 
      Model  & Density Profile$^{\dagger}$ $\rho_{\rm i}$ & 
	Critical Temperature $T^{*}$\\ \hline
      1 & $\propto (1-r^{2})$ &         0.4882608 \\
      2 & $\propto (1-r^{2})^{2}$ &     0.5351886 \\
      3 & $\propto \cos^{2}(\pi r/2)$ & 0.5470739 \\
      4 & $\propto \theta(1-r)$ &       0.2752568 \\
    \end{tabular}
  \end{center}
{\footnotesize $^{\dagger}$ The total mass and initial size of the cloud 
are normalized to be unity.}
\end{table}

\begin{table}[htbp]
	\caption{\label{tb:IMF} $M_{\rm p}$ and $M_{\rm av}$ 
for the formed cores.}
	\begin{center}
		\begin{tabular}{cccc} 
		$(p,q)$ & $M_{\rm av}/M_{\rm tot}$ & $M_{\rm p}/M_{\rm tot}$  \\ \hline\hline
		$(1.0,0.1)$ & 0.87  & 0.94 \\ \hline 
		 $(1.0,0.7)$ & 0.90  & 1.0 \\ \hline
                $(0.1,0.1)$ &  0.86 & 0.95 \\ \hline
                $(0.1,0.7)$ &  0.90 & 1.0 \\

		\end{tabular}
	\end{center}
\end{table}

\newpage

\begin{thebibliography}{99}
\bibitem{chandrasekhar1939}
S.~Chandrasekhar, 
{\it An Introduction to the Study of Stellar Structure} 
(Dover, New York, 1939).
\bibitem{padmanabhan1990}
T.~Padmanabhan,
Phys. Rep. 
{\bf 188}, 285 (1999).
\bibitem{devega1996}
H.J.~de Vega, N.~S\'anchez and F.~Combes,
Nature (London)
{\bf 383}, 53 (1996).
\bibitem{semelin1999}
B.~Semelin, H.J.~de Vega, N.~S\'anchez and F.~Combes, 
Phys. Rev. D 
{\bf 59}, 125021 (1999).
\bibitem{choptuik1993}
M.W.~Choptuik,
Phys. Rev. Lett.
{\bf 70}, 9 (1993).
\bibitem{ec1994}
C.R.~Evans and J.S.~Coleman,
Phys. Rev. Lett.
{\bf 72}, 1782 (1994).
\bibitem{kha1995}
T.~Koike, T.~Hara and S.~Adachi,
Phys. Rev. Lett.
{\bf 74}, 5170 (1995).
\bibitem{maison1996}
D.~Maison,
Phys. Lett. B
{\bf 366}, 82 (1996).
\bibitem{kha1999}
T.~Koike, T.~Hara and S.~Adachi,
Phys. Rev. D 
{\bf 59}, 104008 (1999).
\bibitem{nc2000}
D.W.~Neilsen and M.W.~Choptuik,
Class. Quantum Grav. {\bf 17}, 733 (2000);
{\bf 17}, 761 (2000).
\bibitem{yokoyama1998}
  J.~Yokoyama,
  Phys. Rev. D
  {\bf 58}, 107502 (1998).
\bibitem{nj1998}
  J.C.~Niemeyer and K.~Jedamzik,
  Phys. Rev. Lett.
  {\bf 80}, 5481 (1998).
\bibitem{gundlach1999} 
C.~Gundlach,
Living Rev. Relativ. 
{\bf 2}, 4 (1999).
\bibitem{cc1999}
B.J.~Carr and A.A.~Coley,
Class. Quantum Grav.
{\bf 16}, R31 (1999).
\bibitem{carr1993}
B.J.~Carr,
(unpublished).
\bibitem{penston1969}
M.V.~Penston,
Mon. Not. R. Astron. Soc. {\bf 144}, 425 (1969).
\bibitem{larson1969}
R.B.~Larson,
Mon. Not. R. Astron. Soc.
{\bf 145}, 271 (1969).
\bibitem{hn1997}
T.~Hanawa and K.~Nakayama,
Astrophys. J. 
{\bf 484}, 238 (1997).
\bibitem{hm2000}
T.~Hanawa and T.~Matsumoto,
Publ. Astron. Soc. Jpn.
{\bf 52}, 241 (2000);
Astrophys. J. 
{\bf 521}, 703 (2000).
\bibitem{ti1999}
T.~Tsuribe and S.~Inutsuka,
Astrophys. J. {\bf 526}, 307 (1999).
\bibitem{hm2001}
T.~Harada and H.~Maeda,
Phys. Rev. D 
{\bf 63}, 084022 (2001).
\bibitem{op1987}
A.~Ori and T.~Piran,
Phys. Rev. Lett.
{\bf 59}, 2137 (1987);
Gen. Relativ. Grav.
{\bf 20}, 7 (1988).
\bibitem{op1990}
A.~Ori and T.~Piran,
Phys. Rev. D
{\bf 42}, 1068 (1990).
\bibitem{harada1998}
T.~Harada,
Phys. Rev. D 
{\bf 58}, 104015 (1998).
\bibitem{hin2002}
T.~Harada, H.~Iguchi, and K.~Nakao,
Prog. Theor. Phys. 
{\bf 107}, 449 (2002).
\bibitem{harada2001}
T.~Harada,
Class. Quantum Grav. 
{\bf 18}, 4549 (2001).
\bibitem{op1988}
A.~Ori and T.~Piran,
Mon. Not. R. Astron. Soc.  
{\bf 234}, 821 (1988).
\bibitem{mh2001}
H.~Maeda and T.~Harada,
Phys. Rev. D 
{\bf 64}, 124024 (2001). 
\bibitem{hunter1977}
C.~Hunter,
Astrophys. J.
{\bf 218}, 834 (1977).
\bibitem{ps1974}
  W.H.~Press and P.~Schechter,
  Astrophys. J.
  {\bf 187}, 425 (1974).
\bibitem{maeda2002}
H.~Maeda,
(in preparation).
\end{thebibliography}
\end{document}